\documentclass[10pt, twocolumn, aps, prd, nofootinbib, superscriptaddress, preprintnumbers]{revtex4-1}

\pdfoutput=1

\usepackage{amsmath, amssymb, amsfonts}
\usepackage{bm,bbm}
\usepackage{graphicx, graphics}
\usepackage{color}
\usepackage{hyperref}
\usepackage{cleveref}
\usepackage[utf8]{inputenc}
\usepackage{slashed}
\usepackage[normalem]{ulem}
\usepackage{tabularx}
\usepackage{url}

\newcommand{\eq}[1]{(\ref{e.#1})}
\newcommand{\eql}[1]{\label{e.#1}}
\newcommand{\Fig}[1]{Fig.~\ref{f.#1}}

\newcommand{\Sec}[1]{Section~\ref{s.#1}\@}

\newcommand{\beq}{\begin{equation}\begin{aligned}}
\newcommand{\eeq}{\end{aligned}\end{equation}}
\newcommand{\beqa}[1]{\begin{equation}\begin{alignedat}{#1}}
\newcommand{\eeqa}{\end{alignedat}\end{equation}}
\newcommand{\Cases}[1]{\left\{ \begin{aligned} #1 \end{aligned} \right.}

\newcommand{\fr}[2]{\dfrac{#1}{#2}}
\newcommand{\wba}[1]{\overline{#1}}

\newcommand{\tr}{\mathrm{tr}\,}

\newcommand{\diag}{\mathrm{diag}\,}

\renewcommand{\Im}{\mathrm{Im}\,}

\newcommand{\lt}{\!\left}
\newcommand{\rt}{\right}
\newcommand{\bra}{\langle}
\newcommand{\ket}{\rangle}

\newcommand{\lsim}{\lesssim}
\newcommand{\gsim}{\gtrsim}
\newcommand{\dt}{\!\cdot\!}

\newcommand{\del}{\partial}
\newcommand{\dd}{\mathrm{d}}
\newcommand{\DD}{\mathrm{D}}

\newcommand{\I}{\mathrm{i}}
\newcommand{\e}{\mathrm{e}}
\newcommand{\C}{\mathrm{c}}
\newcommand{\T}{\mathrm{T}}

\newcommand{\al}{\alpha}

\newcommand{\ga}{\gamma}
\newcommand{\Ga}{\Gamma}
\newcommand{\de}{\delta}
\newcommand{\ep}{\epsilon}
\newcommand{\ka}{\kappa}
\newcommand{\la}{\lambda}
\newcommand{\La}{\Lambda}
\newcommand{\sg}{\sigma}
\newcommand{\sgb}{{\bar{\sigma}}}

\newcommand{\cL}{{\mathcal L}}
\newcommand{\cM}{{\mathcal M}}
\newcommand{\cO}{{\mathcal O}}

\newcommand{\ev}{\>\text{eV}}
\newcommand{\Mev}{\>\text{MeV}}
\newcommand{\gev}{\>\text{GeV}}
\newcommand{\tev}{\>\text{TeV}}
\newcommand{\fb}{\>\mathrm{fb}}
\newcommand{\micron}{\>\mu\mathrm{m}}

\newcommand{\als}{\al_\mathrm{s}}

\newcommand{\MET}{\slashed{E}_\T}

\newcommand{\SU}{\mathrm{SU}}

\newcommand{\U}{\mathrm{U}}
\newcommand{\Z}{\mathbbm{Z}}

\newcommand{\cc}{\mathrm{c.c.}}

\newcommand{\PS}{{\phantom.}}

\newcommand{\F}{{\text{\tiny F}}}
\newcommand{\s}{{\text{\tiny S}}}
\newcommand{\EQ}{\text{eq}}
\newcommand{\epCP}{\ep_\text{\tiny CP}}

\begin{document}

\title{LHC Signatures of WIMP-triggered Baryogenesis}

\author{Yanou Cui}
\affiliation{Department of Physics and Astronomy, University of California-Riverside, Riverside, CA 92521, USA}
\affiliation{Perimeter Institute, 31 Caroline Street, North Waterloo, Ontario N2L 2Y5, Canada}
\affiliation{Maryland Center for Fundamental Physics, University of Maryland, College Park, MD 20742, USA}

\author{Takemichi Okui}
\affiliation{Department of Physics, Florida State University, Tallahassee, FL 32306, USA}

\author{Arash Yunesi}
\affiliation{Department of Physics, Florida State University, Tallahassee, FL 32306, USA}

\begin{abstract}
A robust mechanism was recently proposed in which thermal freeze-out of WIMPs can provide a unified origin of dark matter and baryon abundances in our universe. We point out that this WIMP-triggered baryogenesis mechanism can exhibit a rich collider phenomenology and be tested at the current and near-future experiments at LHC, even in the case where the WIMPs are completely devoid of SM gauge and higgs portal interactions, as may be motivated by the persistent null results of WIMP dark matter searches. We catalogue a rich array of LHC signatures robustly present in such a scenario. In particular, the simplest such implementation can already offer a very clean signal of a TeV-scale resonance that decays to diphotons with a cross section that can easily be within the reach of the current and near-future LHC runs in the region of parameter space that leads to a successful baryogenesis. Other characteristic signatures include the production of multi-bottom and/or multi-top quarks, promptly or displaced. An even more exotic possibility is the production of two separate sets of isolated emerging jets connected by a charged track, which may require new dedicated studies. Finally, di-nucleon decay can also provide a powerful probe of the mechanism.
\end{abstract}

\maketitle

\section{Introduction}
\label{s.intro}
The origins of dark matter (DM) and the large matter-antimatter asymmetry in the present universe are two of the biggest mysteries in fundamental physics. 
For DM, an attractive scenario is provided by the ``thermal WIMP freeze-out'' paradigm, 
where a quick estimate shows that the present-day abundance of stable particles of a weak-scale mass would roughly agree with observation if the particles were once in equilibrium with the particles of the Standard Model (SM) and then decoupled from the SM bath as they underwent annihilation into some lighter particles via an interaction of weak-force strength \cite{Lee:1977ua}. 
The existence of such weakly interacting massive particles (WIMPs) is an attractive possibility as it only assumes a mass scale that is already known to exist and is being actively explored at the LHC, and an interaction strength that is comfortably perturbative but not unnaturally small. 
Moreover, a variety of theories addressing the electroweak hierarchy problem explicitly predict WIMPs with potentially diverse lifetimes. 
Much attention has also recently been drawn to the apparent coincidence of the baryon and DM abundances, $\Omega_\text{B} \sim \Omega_\text{DM}$, which may be suggesting a common origin for cosmic baryons and DM. 
With all these motivations, it is quite desirable to construct a mechanism of baryogenesis in which a WIMP and its thermal freeze out play a crucial role.

The WIMP-triggered baryogenesis scenario proposed by Cui and Sundrum~\cite{Cui:2012jh} provides a robust such mechanism.
(For other baryogenesis mechanisms using thermal WIMPs, see Refs.~\cite{McDonald:2011zza, Cui:2011ab, Davidson:2012fn}.)
The idea is neat and simple. 
Consider a WIMP (different from a DM WIMP) that is meta-stable and decays to SM quarks in a baryon-asymmetric manner, with a lifetime so long that the decay occurs well after its freeze-out.
Being a WIMP, the meta-stable WIMP has a freeze-out abundance similar to that of the DM WIMP
and thereby leads to $\Omega_\text{B} \sim \Omega_\text{DM}$---nicely in agreement with observation---if we assume an $\cO(1)$ CP violation and ignore the difference between the QCD and weak scales.%
\footnote{It is fortunate that the QCD and weak scale are only a couple of orders of magnitude apart, although this proximity of the two scales is admittedly not explained.}

In addition, the simplest realization of this scenario has a structure that almost calls for the embedding of the model into a supersymmetric (SUSY) extension of the SM with an R-parity violation (RPV)~\cite{Cui:2012jh, Cui:2013bta}. 
(We will also see this in \Sec{lagrangian}.)
Since the production of baryon asymmetry from the decays of meta-stable WIMPs occurs at temperatures much below the weak scale, such supersymmetric theories are safe from the washout of baryon asymmetry by RPV decays/scatterings of squarks~\cite{Cui:2012jh, Cui:2013bta}, unlike those with conventional baryogenesis mechanisms in which baryon asymmetry is generated at much higher scales. 
It can also have a distinct collider phenomenology, in particular in displaced vertex search channels, offering us the opportunity to directly probe this baryogenesis mechanism at the LHC~\cite{Cui:2014twa}.

On the other hand, the growing null results from direct/indirect/collider WIMP searches make it increasingly more likely that WIMPs, if they exist, are completely devoid of SM gauge interactions.  
In the context of DM only, such a ``hidden'' WIMP scenario may have no hope of being experimentally probed except through cosmological measurements such as those of the matter power spectrum and/or cosmic microwave background spectra ~\cite{Chacko:2015noa, Chacko:2016kgg}.
This would especially be the case in the absence of the ``extra'' particles provided by SUSY\@.

In this paper, however, we point out that the prospect for probing a hidden WIMP sector is quite bright in the WIMP-triggered baryogenesis scenario.
Essentially, the reason for this is that a WIMP-triggered baryogenesis mechanism cannot be completely hidden---even if the WIMPs themselves are hidden---because it must somehow connect to baryons to make baryogenesis happen.
We thus expect some inevitable signatures at hadron colliders.
In particular, we will see that the simplest hidden-sector realization of WIMP-triggered baryogenesis contains new colored and electrically charged scalars $\phi$ in addition to WIMPs. 
The baryogenesis mechanism requires $\phi$ to couple to the meta-stable WIMP and a SM quark.   
Hence, $\phi$ can be pair-produced from gluons or from quarks in association with a pair of meta-stable WIMPs.

The purpose of this paper is to highlight a variety of experimental signatures involving $\phi$, the WIMPs, and a mediator $S$ responsible for setting the freeze-out abundances of the WIMPs. These particles are all integral components of the mechanism.
First, in \Sec{baryo}, we present the structure of the simplest hidden-sector realization of WIMP-triggered baryogenesis and work out constraints from the observed baryon abundance in the universe.
In \Sec{phi_pheno}, we discuss an extremely rich array of possible decay modes of $\phi$, including displaced multi-jet and displaced multi-top/bottom quark productions.
A more exotic possibility is the production of two separate sets of isolated emerging jets connected by a charged track.
We also point out that di-nucleon decay is a powerful probe into the physics of $\phi$.
In \Sec{diphoton}, we study the resonance $S$, focusing on its particularly clean decay channel to $\ga\ga$, which is necessarily generated at 1-loop via the loop of $\phi$.
In \Sec{WIMPs}, we look at the direct production of the WIMPs. Since the WIMPs responsible for baryogenesis are not stable, they can decay within the LHC detector if their lifetimes are sufficiently short. The decay products involve $\phi$ and thus inherit the rich $\phi$ phenomenology.
These signals---some of them could appear simultaneously, some other are mutually exclusive---can provide us with nontrivial pieces of information on the structure of the theory of WIMP-triggered baryogenesis such as the mass spectrum and flavor structure.
\Sec{colliderpheno} will cover various cases where such experimental probes may be possible.
In some cases, it may even be possible to make quantitative connections between the collider measurements and the cosmic baryon abundance.

\section{The WIMP-triggered baryogenesis}
\label{s.baryo}
Here we review the WIMP-triggered baryogenesis mechanism proposed in~\cite{Cui:2012jh} and write down a concrete model that realizes the scenario. 
Our model is very similar to the one in~\cite{Cui:2012jh} up to some minor modifications. However, unlike in~\cite{Cui:2012jh} where the purpose of the model is to provide an ``existence proof'' of the mechanism, 
we would like to argue that the model is not merely an example that works but actually is a robust, representative realization of the scenario. Thereby, we wish to provide a strong motivation for its collider signatures as experimental probes for the WIMP-triggered baryogenesis mechanism.

\subsection{The field content and Lagrangian}
\label{s.lagrangian}
We begin by assuming the existence of a meta-stable WIMP $\chi_1^\PS$ (in addition to an absolutely stable DM WIMP $\chi_0^\PS$) that is completely neutral under the SM gauge group.%
\footnote{It should be noted, however, that the mechanism does clearly allow the possibility that DM is not a WIMP so the existence of $\chi_0$ is logically optional, although our philosophy here is to provide a unified story of DM and baryogenesis.}
Since $\chi_1^\PS$ is meta-stable and no longer present in the universe today, 
direct or indirect WIMP detection experiments do not constrain the properties of $\chi_1^\PS$. However,
the null results of those experiments so far makes it increasingly more motivated to consider the case where the DM WIMP, $\chi_0^\PS$, possesses no SM gauge interactions.
Then, it is natural to take the meta-stable WIMP, $\chi_1^\PS$, to be also a SM-gauge singlet, as the underlying philosophy of the WIMP-triggered baryogenesis mechanism is to associate a single framework with both the DM abundance and baryon asymmetry.    
As we will see below, the WIMP sector also needs a third WIMP, $\chi_2^\PS$, in order to  have a CP-violating interference in the $\chi_1^\PS$ decay (as necessary for baryogenesis). It is amusing to note that the WIMP sector consists of three generations of matter just like the SM\@.
   
The possibility of such WIMP sector well shielded from the SM sector is especially robust if the WIMPs are spin-$1/2$ fermions.
If the WIMPs are scalars, the symmetries that allow a WIMP mass term, $\chi^\dag \chi$, would necessarily also allow a renormalizable quartic interaction $\chi^\dag \chi H^\dag \! H$ with the SM higgs doublet $H$.
On the other hand, if the WIMPs are spin-$1/2$ fermions, 
gauge invariance allows only one renormalizable interaction $H\ell \chi$, 
where $\ell$ is an SM lepton doublet. 
But $H\ell \chi$ can easily be forbidden by a global symmetry 
(e.g., a $\Z_2$ under which all SM lepton fields ($\ell$ and $e^\C$) are odd), 
thereby separating the WIMPs from the SM particles completely at the renormalizable level.
If the WIMPs have a spin higher than $1/2$, that could also naturally explain the separation of the WIMPs from the SM sector, but it would come with the whole baggage of a symmetry breaking sector to give mass to the WIMPs. Therefore, as in \cite{Cui:2012jh}, 
we consider the minimal possibility that the WIMPs are spin-$1/2$, SM-gauge neutral, Majorana fermions described by
\beq
\cL_\chi = 
\I \chi_i^\dag \sgb \dt \del \chi_i^{\phantom{\dag}} 
-\fr{m_{\chi_i^\PS}}{2} (\chi_i^\PS \chi_i^\PS + \cc)
\eeq
with three SM-gauge singlet, 2-component spinors $\chi_{0,1,2}^\PS$.

Next, we need to introduce a mediator particle $S$ through which the WIMPs annihilate into lighter particles before their number density freezes out. 
Since it couples to a pair of the SM-gauge singlet fermionic WIMPs, $S$ has to be a SM-gauge singlet boson. Letting $S$ carry spin-1 or higher would require an additional higgs sector to give mass to $S$, thereby significantly complicating the model for no reason.
Therefore, the simplest possibility is that $S$ is a real scalar described by
\beq
\cL_S = \fr12 (\del S)^2 - \fr{m_S^2}{2} S^2 
\,.
\eeq
As we will see below, $S$ automatically acquires couplings to $gg$ and $\ga\ga$ at 1-loop, 
offering a particularly clean di-photon signal to be searched for at the LHC\@.

After the annihilation process through $S$ freezes out,
$\chi_1^\PS$ must decay in a way that violates CP and baryon number.
In order to maintain the attractive assumptions behind the thermal WIMP freeze-out framework, we do not wish to introduce any mass scale other than the weak scale. 
Our lagrangian should thus only contain weak-scale mass terms and dimensionless gauge/Yukawa couplings. 
Hence,  $\chi_1^\PS$ must couple to a fermion and a boson. 
This fermion has to be an SM quark to introduce the (violation of) baryon number to the story.%
\footnote{\label{ftnote:leptogenesis}%
Leptogenesis is also a possibility in principle but then the leptogenesis would have to be complete before the electroweak sphaleron ceases to be active, 
which would thus require the $\chi_1^\PS$ decay temperature to be above the weak scale and hence the freeze out temperature $T_\F$ even higher. 
We can prevent $m_{\chi_1}$ from being even more far away from the weak scale by 
having the mediator $S$ sufficiently heavy and/or have small couplings so that $T_\F \sim m_{\chi_1}$ \cite{Cui:2013bta}, but this would make $S$ inaccessible at the LHC\@.
In this paper, we thus focus on a ``direct'' baryogenesis scenario, 
which can be realized with weak-scale $m_S$ and $m_{\chi_1}$.}
Then, since $\chi_1^\PS$ is a SM-gauge singlet, 
the boson has to carry the same gauge charges as the SM quark. 
It thus has to be a new particle, which we call $\phi$.
The boson $\phi$ should then subsequently decay to a pair of SM quarks such that
the three SM quarks coming out at the end of the $\chi_1^\PS$ decay chain have a nonzero net baryon number.
We assume $\phi$ is a scalar, again because otherwise we would need an additional higgs sector to generate its mass. 
However, the existence of a new particle with a weak-scale mass that couples to a quark is dangerous as it would generically induce excessive flavor violating processes in the quark sector. As we will discuss later, the simplest symmetry solution to this problem is to have three generations of $\phi$, i.e., we have
\beq
\cL_\phi = |\DD \phi_i|^2 - m_{\phi_i}^2 \phi_i^{\dag} \phi_i^\PS 
\eeq
with $i=1,2,3$ and, for definiteness, we take the gauge charges of $\phi$ to be $(\mathbf{3}^*, \mathbf{1})_{-2/3}$, i.e.,
the same as the right-handed up-type squark in a supersymmetric theory.
This choice is not unique and we will make comment on other choices later whenever it is possible to do so without too much digression.
It is interesting to note that we again have three generations of matter fields.
It is also intriguing that the existence of three generations of $\phi$ is readily compatible with a supersymmetric embedding of our model where the three $\phi$ scalars are literally the three right-handed up-type squarks. See Refs.~\cite{Cui:2012jh, Cui:2013bta} for further supersymmetric explorations of the scenario. 

Now, we are ready to write down the interactions that are essential to our discussions (non-essential interactions will be discussed later): 
\beqa{2}
\cL_\text{int} 
&=\,&& 
-\fr12 S (y_{i}^\PS \chi_i^\PS \chi_i^\PS + \cc) 
-(\la_{i} \, \phi^\dag u^\C \chi_i^\PS + \cc)
\\& &&
-\ka \, S \phi^\dag \phi
-(\ga \phi d^\C d^\C + \cc)
\,,\eql{interactions}
\eeqa
where $u^\C$ and $d^\C$ are the up- and down-type anti-quark fields of the SM\@.
The gauge and flavor indices are implicit except for the $\chi$ flavor, $\chi_i^\PS$ ($i=0,1,2$).
Without the $\la_i$ couplings, the Lagrangian would have three $\Z_2$ symmetries, $\Z_2^{(i)}$ ($i=0,1,2$) under which $\chi_i^\PS$ is odd and everything else even.
Since $\chi_0^\PS$ has to be stable to constitute DM, 
we assume $\Z_2^{(0)}$ is exact and hence $\lambda_0 = 0$ exactly.
On the other hand, we want $\chi_1^\PS$ to be meta-stable, 
so we assume that $\Z_2^{(1)}$ is slightly broken by a tiny, nonzero value of $\la_1$. 
We do not assume $\Z_2^{(2)}$ at all as   
there is no need for $\chi_2^\PS$ to be stable or meta-stable.
The $y_{0,1}^\PS$ couplings are (partly) responsible for setting the abundances of $\chi_{0,1}^\PS$ (before $\chi_1^\PS$ decays).
The decay rate of $\chi_1^\PS$ is given at the tree level by
\beq
\Ga_{\chi_1^\PS} 
= \fr{9 \, |\la_1|^2 \, m_{\chi_1^\PS} }{16\pi} 
  \lt( 1 - \fr{m_\phi^2}{m_{\chi_1^\PS}^2} \rt)^{\!\! 2},
\eql{Gamma_chi1}
\eeq
where the $9$ is due to the 3 colors and 3 generations of $u^\C$ and $\phi$.%
\footnote{We have evidently assumed that the flavor of $\phi$ is perfectly correlated with that of $u^\C$. We have also assumed that $\la_1$ is flavor independent. Justifications of these assumptions will be discussed around \eq{MFVrelations}.}
The mass of $u^\C$ has been neglected for simplicity.
In order for $\chi_1$ to decay well after its freeze-out and well before big-bang nucleosynthesis (BBN), we must demand that
\beq
\sqrt{g_{*\text{\tiny BBN}}} \fr{T_\text{\tiny BBN}^2}{M_*} \ll \Ga_{\chi_1^\PS} \ll \sqrt{g_{*\F}} \fr{T_\F^2}{M_*}
\,,\eql{lambda1window}
\eeq
where $g_{*\text{\tiny BBN}} = 10.75$ is the effective number of relativistic degrees of freedom right before BBN begins at $T = T_\text{\tiny BBN} \sim 1\Mev$, while $g_{*\F}$ is its counterpart at the time of $\chi_1^\PS$ freeze-out at $T=T_\F$. The scale $M_*$ is the combination of numbers that frequently appears in cosmology:
\beq
M_* \equiv \lt( \fr{8\pi G_\text{N}}{3}  \fr{\pi^2}{30} \rt)^{\!\! -\frac12} \!
\simeq \fr{M_\text{P}}{1.66}
\,,\eql{M*}
\eeq
where $M_\text{P}$ denotes the Planck mass, $1.22 \times 10^{19}\gev$.
From~\eq{Gamma_chi1} and~\eq{lambda1window}, we clearly see that we must have $|\la_1| \ll 1$, 
which we attribute to a weakly broken $\Z_2$ as we discussed above. 
The allowed window~\eq{lambda1window} is comfortably wide; it is about six-orders-of-magnitude wide since, as $T_\F$ turns out to be about $1/20$ of $m_{\chi_1^\PS} \sim 1\tev$ for a typical WIMP.

As mentioned above, $\la_2$ needs not be small and we assume $|\la_2| \sim \cO(1)$.
(It is amusing to note that the three generations of $\chi$ fermions with hierarchical $\la_{0,1,2}$ are reminiscent of the SM fermions with hierarchical Yukawa couplings.)
Most importantly,
the phase in the product $\la_1^* \la_2^\PS$ cannot be removed by field redefinition,
thereby providing a source of CP violation necessary for baryogenesis
(which is the sole reason for the existence of $\chi_2^\PS$).
Ignoring the mass of $u^\C$ for simplicity, 
the fraction of CP violation $\epCP$ is given at the one-loop level by
\beq
\epCP 
&\equiv 
\fr{\Ga_{\chi_1 \to \phi+\wba{u}^\C} - \Ga_{\chi_1 \to \phi^* + u^\C}}
   {\Ga_{\chi_1 \to \phi+\wba{u}^\C} + \Ga_{\chi_1 \to \phi^* + u^\C}}
\\
&=
\fr{1}{8\pi} 
\fr{\Im\!\bigl[ (\la_1^* \la_2^\PS)^2 \bigr]}{|\la_1|^2} 
\sqrt{x_1 x_2} \,
\bigl( f(x_1, x_2) + g(x_1, x_2) \bigr)
\,,\eql{epCP}
\eeq
where $x_i \equiv m_{\chi_i^\PS}^2 / m_\phi^2$ and 
\beq
& f(x_1, x_2) = \fr{1}{2(x_1 - x_2)} \lt( 1 - \fr{1}{x_1} \rt)^{\!\! 2} ,
\\
& g(x_1, x_2) 
\\
&= 
\Cases{
&\fr{1}{x_1} - \fr{x_1 + x_2 - 2}{(x_1 - 1)^2} \log\fr{x_1 (x_1 + x_2 - 2)}{x_1 x_2 - 1}
&&\>\text{if $x_2 > 1$,}
\\
&\fr{x_1 x_2 - 1}{x_1 (x_1 - 1)} - \fr{x_1 + x_2 - 2}{(x_1 - 1)^2} \log x_1
&&\>\text{if $x_2 < 1$.}
}
\eeq
The function $f(x_1,x_2)$ comes from a self-energy diagram while $g(x_1,x_2)$ comes from a vertex correction diagram.
Needless to say, we have $x_1 > 1$ so that $\chi_1^\PS$ can decay to $\phi$.
We have assumed that $m_{\chi_1^\PS}$ and $m_{\chi_2^\PS}$ are not similar  
so that we never hit the singularity in $f(x_1,x_2)$.%
\footnote{We do not consider the fine-tuned case $m_{\chi_1^\PS} \simeq m_{\chi_2^\PS}$ that could resonantly enhance $\epCP$ as in ``resonant leptogenesis''~\cite{Pilaftsis:2003gt}. 
Since $|\la_1| \ll |\la_2| \sim 1$, such degeneracy would not even be stable under renormalization group running.
Any attempt to justify $m_{\chi_1^\PS} \simeq m_{\chi_2^\PS}$ by a symmetry would have to confront the breaking of that symmetry by $|\la_1| \ll |\la_2|$.}  
Notice that the small magnitude of $\la_1$ (required by the metastability of $\chi_1^\PS$) does not affect the size of $\epCP$ at all. 
Thus, with an $\cO(1)$ phase of $\la_1^* \la_2^\PS$
and an $\cO(1)$ magnitude of $\la_2$, 
and with all the masses around the weak scale,
we see that a ``typical'' size of $\epCP$ is $\cO(10^{-2})$. 

The $\ka$ coupling in~\eq{interactions} cannot be forbidden by any symmetry that allows the $y_1$ coupling. The $y_1$ coupling is necessary for mediating the annihilation of $\chi_1$.
One might think that the $\la_1$ coupling could also mediate the annihilation of $\chi_1$ into $u^\C$ via $t$-channel $\phi$ exchange, which would be more economical because it would not need $S$.  
However, the condition~\eq{lambda1window} forces $\la_1$ to be too small to give rise to a large enough annihilation rate.
We therefore need the mediator $S$ and coupling $y_1$.
Once $y_1$ is introduced, no symmetry can forbid $\ka$.
One annihilation channel of $\chi_1$ is then given by $\chi_1\chi_1 \to \phi\phi^*$ via an $s$-channel $S$ involving both the $y_1$ and $\ka$ couplings.
We also see that the $\ka$ coupling necessarily gives rise to $S \to gg$ and $S \to \ga\ga$ via a loop of $\phi$, predicting the existence of a clean di-photon signal to be searched for at the LHC\@.
Since $\ka$ is a relevant coupling allowed by symmetry, its significant presence does not require any further justification as its effects grow at low energies. Rather, we must make sure that it is not too large. In particular, the strong attractive force between $\phi$ particles mediated by $S$ exchange should not cause $\phi$ to condense in the vacuum and spontaneously break the $\SU(3)$ color and $\U(1)$ electromagnetism gauge symmetries. Noting that $\phi$ has 3 colors and comes in 3 generations, such strong coupling limit would correspond to $\ka^2 / m_S^2 \lsim 16\pi^2 / 3^2$, i.e., 
\beq
\fr{\ka}{m_S} \lsim \fr{4\pi}{3}
\,.\eql{kappa_bound}
\eeq
We adopt this as our theoretical upper bound on $\ka$.
(Without loss of generality, we have taken $\ka$ to be positive by absorbing its sign into $S$.)

The $\ga$ coupling in~\eq{interactions} is responsible for $\phi$ decay,
which quickly converts the CP asymmetry~\eq{epCP} into a baryon asymmetry.
For example, the decay chain $\chi_1^\PS \to \phi + \wba{u}^\C \to \wba{d}^\C + \wba{d}^\C + \wba{u}^\C$ increases the baryon number by one. 
It is important to do this in ``two steps'' with an on-shell intermediate $\phi$,
because the imaginary part in~\eq{epCP} arises from a region of the loop momentum space where the $\phi$ in the loop goes on-shell.
If we instead only have an off-shell $\phi$, a CP asymmetry could still be generated but only as an interference between tree and 1-loop diagrams for a \emph{3-body} decay.
That would lead to much smaller $\epCP$ and make baryogenesis much harder, although it is possible \cite{Cui:2013bta, Monteux:2014hua}.    
We thus assume that $\chi_1^\PS$ can decay to $\phi$, i.e., 
\beq
m_\phi < m_{\chi_1^\PS}
\,.\eql{mphi:upperbound}
\eeq
For various checks ensuring the baryon asymmetry thus produced not to be washed out, see~\cite{Cui:2012jh}.

Each of $\la_1$, $\la_2$, and $\ga$ has implicit quark and $\phi$ flavor indices. 
Since all observations of quark flavor violation so far are consistent with SM predictions, any flavor-dependent new physics at the weak scale must have a flavor structure quite akin to the flavor structure of the SM\@. 
That means that the new physics must respect, to a very good approximation, the property of the SM that the $\SU(3)_q \times \SU(3)_u \times \SU(3)_d$ flavor symmetry is only violated by the $Y_u$ and $Y_d$ Yukawa coupling matrices.
In other words, minimal flavor violation (MFV)~\cite{Chivukula:1987py, Hall:1990ac, Buras:2000dm, D'Ambrosio:2002ex} must hold to a good approximation. 
The minimal way to incorporate MFV in the $\la_{1,2}$ and $\ga$ couplings is to have three generations of $\phi$ that form a flavor multiplet transforming like $u^\C$ under the $\SU(3)^3$ quark flavor symmetry of the SM\@.
MFV then dictates that the leading flavor structures should be given (at least to a good approximation) by 
\beq
m_\phi^2 \propto \mathbbm{1}
\,,\quad
\la_{1,2} \propto \mathbbm{1}
\,,\quad
\kappa \propto \mathbbm{1}
\,,\quad
\ga \propto \ep Y_u Y_d Y_d
\,,\eql{MFVrelations}
\eeq
where $\mathbbm{1}$ is an identity matrix in the $\SU(3)_u$ flavor space, 
while $\ep$ is a 3d Levi-Civita tensor that contracts the three $\SU(3)_q$ indices from  $Y_u Y_d Y_d$.
So, $\ga$ has three implicit flavor indices, i.e.,  
one $\SU(3)_u$ index from $Y_u$ and two $\SU(3)_d$ indices from $Y_d Y_d$.
In the $\ga \phi d^\C d^\C$ interaction, those indices of $\ga$ are contracted with the one $\SU(3)_u$ and two $\SU(3)_d$ indices of $\phi$ and $d^\C d^\C$, respectively. 
Writing the flavor indices explicitly, we thus have
\beq
\cL_\text{int} \supset
\ga \phi d^\C d^\C
=c \ep^{ijk} (Y_u)_i^{~a} (Y_d)_j^{~p} (Y_d)_k^{~q} \, \phi_a^\PS d^\C_p d^\C_q
\,,\eql{basis_free}
\eeq
where $c$ is an overall multiplicative free parameter, $i,j,k$ are $\SU(3)_q$ indices, $a$ is an $\SU(3)_u$ index, and $p,q$ are $\SU(3)_d$ indices. 
Without loss of generality, we can go to a basis where
$Y_d = y_d \equiv \diag(m_d/m_b, m_s/m_b, 1)$ and 
$Y_u = V^\dag y_u \equiv V^\dag \, \diag(m_u/m_t, m_c/m_t, 1)$,
where $V$ is the CKM matrix.
Then, the $\ga$ couplings become
\beq
\ga \phi d^\C d^\C
=c \ep^{ijk} V^*_{ai} \, 
(y_u)_a^\PS (y_d)_j^\PS (y_d)_k^\PS \, 
\phi_a^\PS  d^\C_j d^\C_k
\,.\eql{basis_fixed}
\eeq
Since $c$ is a free parameter, 
there is no loss of generality in our normalization conventions, $(y_u)_{33} = (y_d)_{33} = 1$.

The existence of three generations of $\phi$ and their MFV couplings%
\footnote{\label{ftnote:decoupled_phi_12} In principle, it is possible to decouple $\phi_{1,2}$ (i.e., the right-handed sup and scharm) while keeping $\phi_3$ (the right-handed stop) light 
by including a formally higher order term in the MFV expansion as $m_\phi^2 \propto \mathbbm{1} - a Y_u^\dag Y_u$, 
where $a$ is an $\cO(1)$ coefficient chosen such that $(\mathbbm{1} - a Y_u^\dag Y_u)_{33} \ll 1$. 
We do not consider this possibility, 
not only because it is fine tuned in the absence of an underlying model that predicts the value of $a$, 
but also because such decoupling is not sufficient to solve the supersymmetric flavor problem anyway~\cite{ArkaniHamed:1997ab}.}  
clearly indicate that a supersymmetric extension of our model would be that of the ``R-parity violating MFV SUSY''~\cite{Csaki:2011ge} augmented by the WIMPs and $S$.
In such supersymmetric extension, the normalization convention $(y_u)_{33} = (y_d)_{33} = 1$ would actually be realized in the large $\tan\beta$ limit.

Finally, a complete list of all other renormalizable operators allowed by gauge symmetry and $\Z_2^{(0)}$ is 
\beq
& S \,,\quad
S^3 \,,\quad
S^4 \,,\quad
\chi_1^\dag \bar{\sg} \!\cdot\! \del \chi_2^\PS \,,\quad
\chi_1^\PS \chi_2^\PS \,,\quad
S \chi_1^\PS \chi_2^\PS \,,
\\
& S^2 \phi^\dag \phi \,,\quad
S H^\dag H \,,\quad
S^2 H^\dag H \,,\quad
\phi^\dag \phi H^\dag H 
\,.\eql{extras}
\eeq
We adjust the coefficient of the $S$ term such that $\bra S \ket = 0$.
This does not cause any loss of generality, 
because we already have all the couplings (e.g., $\chi\chi$, $S\phi^\dag \phi$) that can be redefined to absorb a nonzero $\bra S \ket$.
The three couplings that mix $\chi_1^\PS$ and $\chi_2^\PS$ can be naturally tiny as they are odd under $\Z_2^{(1)}$, 
which is only slightly broken as required by the meta-stability of $\chi_1^\PS$.
For example, 
if we have none of those three couplings at tree level but have a tiny, nonzero $\la_1$ and an $\cO(1)$ $\la_2$ (because we absolutely need them), 
then the coefficient of $\chi_1^\dag \bar{\sg} \cdot \del \chi_2^\PS$ generated at 1-loop would be of order $\sim 9\la_1 \la_2 / 16\pi^2$, which is minuscule and leads to no consequences worthy of further consideration. 

The $S H^\dag H$ coupling in~\eq{extras} leads to the mixing of $S$ with the SM higgs boson $h$ after electroweak symmetry breaking.
If this mixing is sizable, it can mediate direct annihilation of $\chi_1^\PS$ into SM state.
This is phenomenologically a viable possibility and it was already considered in~\cite{Cui:2012jh}.
Here, adhering to the picture of a hidden WIMP sector, we consider the case where the $S H^\dag H$ coupling is small and does not play a relevant role.
The consistency of this assumption can be seen by setting those couplings to zero at tree level and seeing how large their counter-terms need to be at loop level, 
using the couplings that are already established to exist.
The largest contribution to the $S H^\dag H$ counter-term would come from a two-loop diagram 
where an $S$ becomes a pair of $\chi_2^\PS$ that become a pair of $t^\C$ by exchanging a $\phi$, and then the $t^\C$ pair becomes an $H$ pair by exchanging a $q_3^\PS$. 
Renormalization at a scale of $\cO(1)$ TeV thus requires an $S H^\dag H$ counter-term of order $\sim 3y_2 |\lambda_2|^2 |y_t|^2 m_{\chi_2^\PS} / (16\pi^2)^2 \sim 10^{-4}\tev$. 
This then implies that a natural size of the $S$-higgs mixing angle is given by $10^{-4}\tev \, v / m_S^2$ 
where $v \simeq 246\gev$ is the SM higgs vacuum expectation value. 
Hence, the annihilation of $\chi_1^\PS$ into SM particles via the $h$-$S$ mixing induced by $S H^\dag H$ is highly suppressed.
We thus ignore the $S H^\dag H$ coupling hereafter.
Such reasoning also implies that the $S^3$ coupling can naturally be loop-suppressed. The $S^4$, $S^2 \phi^\dag \phi$, $S^2 H^\dag H$, and $\phi^\dag \phi H^\dag H$ couplings are simply irrelevant for our analyses below.

\subsection{The baryon abundance}
\label{s.abundance}
Let us find the region in the parameter space that gives the observed baryon asymmetry. 
We start with a general formulation without recourse to any specific annihilation channels, and then identify the range of viable parameter space of our model.
By assumption, long before the $\chi_1^\PS$ particles begin to decay, 
they go through a standard WIMP thermal freeze-out process, resulting in a ``would-be'' relic of $\chi_1^\PS$.
Denoting the $\chi_1^\PS$-$\chi_1^\PS$ annihilation cross section by $\sigma$ and the relative speed between the annihilating $\chi_1$'s by $v$,
the freeze-out temperature $T_\F$ can be estimated from 
the instantaneous freeze-out approximation, i.e., 
\beq
\bra \sg v \ket_\F \, n_\F 
\simeq H_\F 
= \sqrt{g_{*\F}} \, \fr{T_\F^2}{M_*}
\,,\eql{freezeout}
\eeq
where $\bra \sg v \ket_\F \equiv \bra \sg v \ket\bigr|_{T = T_\F}$, 
$H_\F = H\bigr|_{T = T_\F}$, 
$n_\F$ is the $\chi_1^\PS$ number density at the freeze-out, 
$g_{*\F} \sim 100$ is the effective number of relativistic degrees of freedom at the time of freeze out,
and $M_*$ is defined in~\eq{M*}.
Then, the $\chi_1^\PS$'s decay with the CP asymmetry $\epCP$ and generate a baryon asymmetry as described in \Sec{lagrangian}.    
The mass density $\rho_\text{\tiny B}^\EQ$ of baryons at the time of matter-radiation equality is thus given by
\beq
\rho_\text{\tiny B}^{\EQ} = m_\text{p} \epCP n_\F \fr{a_\F^3}{a_\EQ^3}
\,,
\eeq
where $m_\text{p}$ is the proton mass (neglecting the tiny difference between the proton and neutron masses), 
and $a_\F$ and $a_\EQ$ are the scale factors of the universe at the freeze-out and the matter-radiation equality, respectively. 
The ratio $a_\F^3 / a_\EQ^3$ is then equal to $g_{*\s,\EQ} T_\EQ^3 / g_{*\F} T_\F^3$ by co-moving entropy conservation,
where $g_{*\s,\EQ} = 2 + \frac78 \cdot 3 \cdot 2 \cdot \frac{4}{11}$.
Then, since the sum of baryon and DM mass densities at the matter-radiation equality must by definition be equal to the photon+neutrino energy density, we obtain 
\beq
(1 + R) m_\text{p} \epCP n_\F 
\fr{g_{*\s,\EQ} T_\EQ^3}{g_{*\F} T_\F^3} 
= \fr{\pi^2}{30} g_{*\EQ} T_\EQ^4
\,,\eql{matter=radiation}
\eeq
where $R \simeq 5.4$ is the observed mass density ratio of DM to baryon abundance, 
$g_{*\EQ} = 2 + \frac78 \cdot 3 \cdot 2 \cdot \lt( \frac{4}{11} \rt)^{\! 4/3}$.
Combining the condition~\eq{matter=radiation} with~\eq{freezeout} to eliminate $n_\F$, we find
\beq
\bra \sg v \ket_\F 
\simeq
(1+R) \epCP 
\fr{m_\text{p}}{m_{\chi_1^\PS}}
\cdot
\fr{30}{\pi^2} 
\fr{g_{*\text{\tiny S},\EQ}}{g_{*\EQ} \sqrt{g_{*\F}}}
\fr{m_{\chi_1^\PS}}{T_\F}
\fr{1}{M_*T_\EQ}
\,,\eql{xsec}
\eeq
where we see that, compared to the standard thermal WIMP DM cross section (i.e., the expression after the ``$\,\cdot\,$'' in \eq{xsec}), 
we need a smaller cross section by the factor of $\epCP m_\text{p} / m_{\chi_1^\PS}$ 
so that we get an over-abundance of $\chi_1^\PS$ to counter the suppressions due to $\epCP$ and $m_\text{p} / m_{\chi_1^\PS}$.
The above expression still contains one unknown ratio, $m_{\chi_1^\PS} / T_\F$.
To obtain this ratio, we substitute the thermal equilibrium density for $n_\F$ in~\eq{matter=radiation}, which gives
\beq
\fr{m_{\chi_1^\PS}}{T_\F} 
&\simeq 
\log r + \fr32 \log(\log r)
\,,\\
r
&\equiv  \fr{2}{(2\pi)^{3/2}} \fr{30}{\pi^2} 
\fr{g_{*\s,\EQ}}{g_{*\EQ} g_{*\F}} 
(1+R) \epCP \fr{m_\text{p}}{T_\EQ}
\,,\eql{TF}
\eeq
where higher order terms involving more logarithms have been neglected.
The required over-abundance of $\chi_1^\PS$ compared to a thermal WIMP DM abundance should mean a higher freeze-out temperature than the DM case.
Indeed, using $R=5.4$ and $T_\EQ = 0.79\ev$ from observations~\cite{Ade:2015xua}, 
the expression~\eq{TF} gives $m_{\chi_1^\PS} / T_\F \simeq 17$ for $\epCP=10^{-2}$ and $g_{*\F} = 100$, which should be compared to $m_{\chi_1^\PS} / T_\F \sim 27$ for a thermal WIMP DM\@.
Combining~\eq{xsec} and~\eq{TF}, we get 
\beq
\bra \sg v \ket_\F
&\simeq 
6 \times 10^{-8} \tev^{-2} \cdot \fr{\epCP}{10^{-2}} \fr{\tev}{m_{\chi_1^\PS}} \fr{10}{\sqrt{g_{*\F}}}
\,,\eql{sigmav:goal}
\eeq
where we have neglected the logarithmic dependences on the ratios of the parameters to their ``benchmark'' values, such as $\log(\epCP / 10^{-2})$.
This is the cross section we need in order to obtain the observed amount of baryon asymmetry.

Now, let us calculate $\bra \sigma v \ket$ in our model.
We have the following three possible annihilation channels, excluding those via higgs-$S$ mixing as it is assumed to be small in our model as we discussed in \Sec{lagrangian}. 
Possibility (i) is $\chi_1^\PS \chi_1^\PS \to \phi\phi^*$ via an $s$-channel $S$ exchange, which is always allowed kinematically because of \eq{mphi:upperbound}.
We can also have 
(ii-a) $\chi_1^\PS \chi_1^\PS \to SS$ with a $t$-channel $\chi_1^\PS$
and (ii-b) $\chi_1^\PS \chi_1^\PS \to SS$ with an $s$-channel $S$, 
if kinematically allowed. 
Finally, we have (iii) $\chi_1^\PS \chi_1^\PS \to \chi_{0,2}^\PS \chi_{0,2}^\PS$ via an $s$-channel $S$ if kinematically allowed.
The amplitude for (i) is $\propto y_1 \ka$, 
those for (ii-a) and (ii-b) are $\propto y_1^2$ and $\propto y_1 \times \text{(the $S^3$ coupling)}$, respectively,  
and that for (iii) goes as $\propto y_1 y_{0,2}$.
All channels can lead to a successful WIMP-triggered baryogenesis. 
However, since the purpose of this paper is to explore collider probes of the WIMP baryogenesis mechanism, we focus on the case where channel (i) dominates.
Since $S$ acquires couplings to $gg$ and $\ga\ga$ via $\phi$ loops, 
the coupling $\ka$ involved in channel (I) can be independently measured at the LHC from $gg \to S \to \ga\ga$. We can also use the process $gg \to S \to \chi_1 \chi_1$ to extract the coupling $y_1$. Then, we will be able to test whether these couplings are indeed responsible for generating the correct freeze-out abundance of $\chi_1$.

It is easy to see that channel (i) is realized in a significant portion of the parameter space.
First, we can simply have $m_{\chi_1} < m_S$ so that (ii-a) and (ii-b) are kinematically forbidden for non-relativistic $\chi_1$. 
Even if they are kinematically allowed, (ii-b) can easily be subdominant to (i) since the $S^3$ coupling can be loop-suppressed as we have discussed in \Sec{lagrangian}.  
Since the amplitudes for (i) and (ii-a) are proportional to $y_1 \ka$ and $y_1^2$, respectively, (i) can also dominate over (ii-a) if, for example, $\ka \sim m_S \sim m_{\chi_1} \sim \cO(1)\tev$ and $|y_1| \ll 1$. 
For (iii), 
the annihilation into $\chi_2^\PS \chi_2^\PS$ can be made subdominant by having $y_2$ small or can simply be kinematically forbidden by assuming $m_{\chi_2} > m_{\chi_1}$.
Similarly, the $\chi_0\chi_0$ channel can be removed if we just assume $m_{\chi_0} > m_{\chi_1}$. 
(Or, $\chi_0$ simply does not exist and DM is unrelated to WIMP-triggered baryogenesis.)
Thus, there certainly exists a large region where the $\chi_1^\PS \chi_1^\PS \to \phi\phi^*$ channel dominates, which is the region we focus on hereafter.%
\footnote{\label{ftnote:decoupling}We do not consider the the fine-tuned possibility that $m_\phi$ is so close to $m_{\chi_1^\PS}$ that the phase space for $\chi_1^\PS \chi_1^\PS \to \phi\phi^*$ is almost closed. 
Not only is it highly tuned but such a case would also generate a highly suppressed $\epCP$ (see~\eq{epCP}), which would in turn require a much larger over-abundance of $\chi_1^\PS$, rendering the whole story less plausible.}

The spin-averaged cross section of $\chi_1^\PS \chi_1^\PS \to \phi\phi^*$ in the nonrelativistic limit of $\chi_1^\PS$, 
away from the resonance region $m_S \approx 2m_{\chi_1^\PS}$,
is given by:
\beq
\sg
= \sg_0 \lt( \fr{\sin^{2\!}\de_1}{v_\chi} + v_\chi \cos^{2\!} \de_1 \rt)
,\eql{xsec:chichi->phiphi}
\eeq
where $v_\chi$ is the speed of $\chi_1\PS$ in the center-of-momentum (CM) frame 
and the $\delta_1$ is the phase in the coupling $y_1$ defined through $y_1 =|y_1| \e^{i\delta_1}$. 
The first and second terms above describe the $s$-wave (from the pseudo-scalar coupling) and $p$-wave (from the scalar coupling) contributions, respectively, as evident from their $v_\chi$ dependences. 
The overall scale $\sg_0$ of the cross section is given by
\beq
\sg_0 =
\fr{9 |y_1|^2\kappa^2}{512\pi m_{\chi_1^\PS}^4} \,
\fr{\sqrt{1 - m_{\phi}^2 / m_{\chi_1^\PS}^2}}
   {\lt(1 - m_S^2 / 4m_{\chi_1^\PS}^2 \rt)^{\! 2}}.
\eql{sg0}
\eeq
where the origin of the $9$ is the same as in~\eq{Gamma_chi1}.
Thermally averaging the cross section~\eq{xsec:chichi->phiphi} then gives
\beq
\bra \sg v \ket_\F
= 2\sg_0  \lt( \sin^{2\!}\de_1 +  \fr{3T_\F}{m_{\chi_1^\PS}} \cos^{2\!} \de_1 \rt).\eql{sigmav}
\eeq
Here $3 T_\F / m_{\chi_1^\PS} \simeq 3/17 \approx 1/6$, so the $p$-wave contribution is not as suppressed as it would be for a thermal WIMP DM\@.
Putting the numbers in, we get
\beq
\bra \sg v \ket_\F
\sim \sg_0 
\simeq 
5 \times 10^{-8} \tev^{-2}
\lt( \fr{|y_1| \ka}{3 \!\times\! 10^{-3} \tev} \rt)^{\!\! 2} \!
\fr{\tev^4}{m_{\chi_1^\PS}^4}
\,,\eql{sigmav:ourprediction}
\eeq
where we have assumed a generic $\cO(1)$ $\de_1$ and also dropped $m_\phi^2$ and $m_S^2$ in~\eq{sg0} for the purpose of estimation.
Equating~the required cross section~\eq{sigmav:goal} and our model's prediction~\eq{sigmav:ourprediction}, we find that the right baryon abundance is generated if
\beq
\lt( \fr{|y_1| \ka}{3 \times 10^{-3} \tev} \rt)^{\!\! 2}
\lt( \fr{1 \tev}{m_{\chi_1^\PS}} \rt)^{\!\! 3}
\sim \fr{\epCP}{10^{-2}} \fr{10}{\sqrt{g_{*\F}}}
\,,\eql{right-abundance}
\eeq
where $\epCP$ is given by~\eq{epCP} and $g_{*\F}$ can be calculated once we determine $T_\F$ from~\eq{TF}.

\section{Collider probes of WIMP-triggered baryogenesis}
\label{s.colliderpheno}

\subsection{The $\phi$}
\label{s.phi_pheno}
As we have seen, the scalar $\phi$ plays an essential role in the WIMP-triggered baryogenesis mechanism.  
Being colored and electrically charged, $\phi$ can be copiously produced at both hadron and lepton colliders,
providing us with experimental probes into the mechanism.

\subsubsection{Di-nucleon decay constraints/signals}
As we mentioned earlier, the $\phi$ scalars can be regarded as the right-handed up-type squarks in the R-parity violating MFV SUSY model~\cite{Csaki:2011ge}.
Ref.~\cite{Csaki:2011ge} performs a comprehensive analysis of constraints on the couplings~\eq{basis_fixed}
from indirect measurements, i.e., those without relying on direct production of $\phi$.
The study concludes that the strongest such constraint comes from di-nucleon decay,
which requires the masses of the right-handed up-type squarks to be $\gsim 400\gev$ for $\tan\beta \sim 40$, 
where the bound depends sensitively on the precise value of poorly known hadronic matrix elements appearing in the di-nucleon decay process. 
Therefore, for $c \sim 1$, the region $m_\phi \gsim 400\gev$ is not excluded.
But this indicates that di-nucleon decay can offer a powerful probe on our scenario in the future if the uncertainties on the hadronic matrix elements are reduced significantly. 

\subsubsection{Collider constraints/signals}
To analyze the collider phenomenology of $\phi$, 
we need to know the dominant interaction in~\eq{basis_fixed} for each of $\phi_{1,2,3}$.
Since $m_d / m_b$ is smaller than the severest suppression $\sim \la^3$ from CKM mixing (where $\la \simeq 0.225$ is the Cabibbo angle), 
it is always better in~\eq{basis_fixed} to have $i=1$ and pay a Cabibbo suppression than having $j=1$ or $k=1$, no matter what $a$ is.
Thus, the dominant terms in the right-hand side of~\eq{basis_fixed} are
\beq
& 2c \fr{m_s}{m_b} \Bigl( 
V^*_{11} \fr{m_u}{m_t} \, \phi_1^\PS 
+ V^*_{21} \fr{m_c}{m_t} \, \phi_2^\PS 
+ V^*_{31} \, \phi_3^\PS 
\Bigr) b^\C s^\C
\\
&= c \bigl( 
6 \times 10^{-7} \phi_1^\PS 
+ 8 \times 10^{-5} \phi_2^\PS 
+ 4 \times 10^{-4} \phi_3^\PS 
\bigr) b^\C s^\C
\,,\eql{phi-bs}
\eeq
where we have extracted relevant measurements from \cite{Agashe:2014kda}. 

At hadron colliders, $\phi$ can easily be pair produced from $gg$ at tree level via QCD interactions. 
The couplings~\eq{phi-bs} show that we can also resonantly produce a $\phi$ from $bs$.
This indeed does occur at the LHC and can lead to an interesting phenomenology~\cite{Kilic:2011sr, Monteux:2016gag}, although 
it is suppressed by the small $\phi b^\C s^\C$ couplings and 
the small $b$-quark PDF\@.
Once (pair-)produced, the subsequent decay chain 
of $\phi$ crucially depends on whether it is lighter or heavier than $\chi_2^\PS$. 
(Recall that $\chi_1$ is heavier than $\phi$ (see Eq.~\eq{mphi:upperbound}).
So, let's look at those two cases separately: 

\underline{{\bf Case 1:} $m_\phi < m_{\chi_2}$}\\
In this case, $\phi$ can only decay to down-type SM quarks through the $\phi d^\C d^\C$ interaction. Then,
the expression~\eq{phi-bs} tells us that all species of $\phi$ dominantly decay to a $b$-jet and a light jet without any $\MET$.
For $c \sim 1$ and $m_\phi \sim 400\gev$, $\phi_2$ and $\phi_3$ clearly decay promptly at collider time scales, 
while the $\phi_1$ decay is barely prompt with a lifetime of $\sim 10^{-13}$~s.
Then, there are two sub-cases of Case 1 depending on whether $\phi$ decays promptly or not:

\textit{\underline{{Case 1-a:} Prompt $\phi$ decay}}\\
Since $\phi$ carries an electric charge, 
it would have been pair-produced at the LEP experiment if sufficiently light, and 
the ALEPH collaboration has placed a limit $m_\phi > 82.5\gev$~\cite{Heister:2002jc}. 
At the Tevatron, the CDF collaboration has excluded the region $50\gev < m_\phi < 125\gev$~\cite{Aaltonen:2013hya}. 
At the LHC, 
the regions $100\gev < m_\phi < 315\gev$ and $200\gev < m_\phi < 385\gev$ have been excluded by the ATLAS~\cite{Aad:2016kww} and CMS~\cite{Khachatryan:2014lpa} collaborations, respectively. 
The most recent ATLAS study~\cite{ATLAS:2016sfd} excludes the regions $250\gev < m_\phi < 405\gev$ and $445 \gev < m_\phi < 510\gev$.
We therefore conclude that, if $m_\phi < m_{\chi_2^\PS}$ and $\phi$ decays promptly, 
the regions $405\gev < m_\phi < 445 \gev$ and $m_\phi > 510\gev$ are currently allowed.

\textit{\underline{{Case 1-b:} Displaced $\phi$ decay}}\\
The bounds are much severer in this case, due to the generally lower SM background for long-lived particle searches.
Ref.~\cite{Liu:2015bma} shows that for the decay length of a few $100\micron$, 
the bound is $m_\phi \gsim 500\gev$. For the decay lengths of order 1~mm to 10~cm, 
the lower bound exceeds $900\gev$.
The bound drops to about $600\gev$ around the decay length of a few m, 
and then again rises to $\sim 900\gev$ once the decay length exceeds the size of the LHC detectors, $\sim 10$~m.

\underline{{\bf Case 2:} $m_\phi > m_{\chi_2}$}\\
In this case, since $\la_2 \sim \cO(1)$ to generate a sizeable CP asymmetry,   
$\phi_{1,2}$ promptly decay as $\phi_{1,2} \to \bar{u}_{1,2} \chi_2^\PS$, i.e.,
\beq
\phi_{1} \to \bar{u} \chi_2^\PS
\,,\quad
\phi_{2} \to \bar{c} \chi_2^\PS
\,.
\eeq
On the other hand, $\phi_3$ promptly decays as
\beqa{3}
\phi_3 &\to \bar{t} \chi_2^\PS
&&\quad\text{if~} 
m_\phi > m_{\chi_2^\PS} \!+ m_t
\,,\\
\phi_3 &\to \bar{b} W^- \chi_2^\PS
&&\quad\text{if~} 
m_{\chi_2^\PS} \!< m_\phi < m_{\chi_2^\PS} \!+ m_t
\,,
\eeqa
where the second case proceeds through an off-shell $\bar{t}$. 
Since $\la_2 \sim \cO(1)$, all of these decays occur promptly.
Subsequently,
the $\chi_2^\PS$ decays as 
\beqa{3}
\chi_2^\PS 
&\to 
tbs \text{~or~} \bar{t} \bar{b} \bar{s}
&&\quad\text{if~} 
m_{\chi_2^\PS} \!> m_t
\,,\\
\chi_2^\PS 
&\to 
cbs \text{~or~} \bar{c} \bar{b} \bar{s}
&&\quad\text{if~} 
m_{\chi_2^\PS} \!< m_t
\,,
\eeqa
where the first case proceeds through an off-shell $\phi_3^\PS$ or $\phi^*_3$
while the second through an off-shell $\phi_2^\PS$ or $\phi_2^*$.
Due to its small coupling~\eq{phi-bs}, $\phi_1$ does not come into play here. 
The $\chi_2^\PS$ decay rate is given by
\beq
\Ga_{\chi_2^\PS} \simeq 
\fr{|\la_2|^2 \xi^2}{512\pi^3} \fr{m_{\chi_2^\PS}^5}{m_\phi^4}
\,,
\eeq
where $\xi = 4\times 10^{-4} c$ if the decay proceeds through an off-shell $\phi_3$,
or $\xi = 8\times 10^{-5} c$ if through an off-shell $\phi_2$.
For simplicity, 
the masses of the final-state fermions as well as higher order terms in $m_{\chi_2} / m_\phi$ have been neglected.
Numerically, the above expression yields
\beq
\Ga_{\chi_2^\PS}^{-1} \sim 
 10^{-13}~\mathrm{s} \cdot
\fr{10^{-8}}{|\la_2|^2 \xi^2}
\lt( \fr{m_\phi}{400\gev} \rt)^{\! 4}
\lt( \fr{200\gev}{m_{\chi_2^\PS}} \rt)^{\!\! 5} 
.\eql{chi2-decay}
\eeq
We thus see that whether $\chi_2^\PS$ is prompt or displaced depends very sensitively on the masses of $\phi$ and $\chi_2^\PS$,
so both cases must be considered:

\textit{\underline{Case 2-a: $m_\phi > m_{\chi_2}$ with well separated $m_\phi$ and $m_{\chi_2}$}}\\
This is the generic case within Case 2.
At colliders, $\phi$ is dominantly pair produced.
Resonant production of a single $\phi$ does occur and is interesting~\cite{Kilic:2011sr, Monteux:2016gag}, but it is suppressed by the small $\phi b^\C s^\C$ couplings~\eq{phi-bs} and 
the small $b$-quark parton distribution function (PDF\@). 
Based on the above discussion on the decay modes of $\phi_i$ and $\chi_2$, 
we find the following full event topology for collider searches:
\begin{itemize}
\item{ 
From $\phi_{1,2}$ pair production we have
\begin{itemize}
\item[(A)]{$jj (jbt) (jb\bar{t})$ or $\ldots t \ldots t$ or $\ldots \bar{t} \ldots \bar{t}$
~if $m_{\chi_2^\PS} \!> m_t$,%
}
\item[(B)]{$jj (jbc) (jbc)$
~if $m_{\chi_2^\PS} \!< m_t$,%
}
\end{itemize}
where $j$ stands for a light jet, $b$ stands for a $b$- or $\bar{b}$-jet, and in light of the recent significant improvement in charm tagging \cite {ATL-PHYS-PUB-2015-001}, we
single out a light charm jet as $c$; the ellipses are used to avoid repetitions.
Each pair of parentheses indicates a displaced vertex if the $\chi_2^\PS$ decay is non-prompt.
So, for example, in the very first case above, the $jj$ is from a primary vertex
and, if $\chi_2^\PS$ is long-lived,  
the $(jbt)$ is from a displaced vertex and the $(jb\bar{t})$ is from another displaced vertex.
}
\item{
From $\phi_{3}$ pair production we have
\begin{itemize}
\item[(C)]{$t\bar{t} (jbt) (jb\bar{t})$ or $\ldots t \ldots t$ or $\ldots \bar{t} \ldots \bar{t}$
~if $m_\phi > m_{\chi_2^\PS} \!+ m_t$ and $m_{\chi_2^\PS} \!> m_t$,%
}
\item[(D)]{$t\bar{t} (jbc) (jbc)$
~if $m_\phi > m_{\chi_2^\PS} \!+ m_t$ and $m_{\chi_2^\PS} \!< m_t$,%
}
\item[(E)]{$bbW^+W^- (jbt) (jb\bar{t})$ or $\ldots t \ldots t$ or $\ldots \bar{t} \ldots \bar{t}$
~if $m_\phi < m_{\chi_2^\PS} \!+ m_t$ and $m_{\chi_2^\PS} \!> m_t$,%
}
\item[(F)]{$bbW^+W^- (jbc) (jbc)$
~if $m_\phi < m_{\chi_2^\PS} \!+ m_t$ and $m_{\chi_2^\PS} \!< m_t$.%
}
\end{itemize}
}
\end{itemize}
Notice that the above event topologies are identical to pair productions of up-type squarks 
followed by each squark decaying to a quark and a neutralino, 
and then the neutralino subsequently decaying to three quarks via an R-parity violating effective 4-fermion interaction.
The bounds clearly depend on the lifetime of $\chi_2^\PS$ (``neutralino'')
that can be prompt, displaced, or collider stable, as well as the Lorentz boost of $\chi_2$ produced from the cascade decay.
Drawing a detailed map of exclusion limits covering all of (A)--(F) clearly requires
a dedicated work of its own and we leave it for future work.

We can, however, already draw some conclusions by noticing that (A)--(F) are similar to the final states considered in various existing SUSY searches.
For example, if $\chi_2^\PS$ is collider stable, 
(A) and (B) are identical to the standard jets$+\MET$ production from two degenerate squark species. 
If $\chi_2^\PS$ decays within the LHC detectors (prompt or displaced), (A) and (B) are similar to gluino pair production followed by the decay of each gluino into quarks and a neutralino that subsequently decays to three quarks via an R-parity violating vertex.
Recasting the limits by the ATLAS and CMS collaborations \cite{CMS:2013jea, Aad:2015lea} for our cases is not straightforward, but there appears no room for $m_\phi$ as light as $\sim 400\gev$. Rather, in many cases, the lower bounds seems well above $\sim 500\gev$ and sometimes reaches $\sim 1\tev$.

\textit{\underline{Case 2-b: $m_\phi > m_{\chi_2}$ with $m_{\phi}\approx m_{\chi_2}$}}\\
This is a special parameter region of Case 2, where $\phi_{1,2}$ barely, but still dominantly and promptly, decay to $\chi_2^\PS$ and $u, c$ quark, while $\phi_3$ decays to $\chi_2^\PS$ and $jjb$ or $\ell \nu b$ via an off-shell $W$ (which itself comes from an off-shell $t$). 
If the subsequent decay of $\chi_2$ is prompt or displaced but still well within the LHC detectors, Case 2-a applies.
However, if $\chi_2$ is collider stable,  
the LHC sensitivity can drop significantly as it relies on hard jets+$\MET$ trigger, 
whereas the jets emitted from $\phi$ decays here would be too soft due to the compressed phase space. 
Indeed, as demonstrated in the SUSY stop searches for such a spectrum~\cite{Aad:2015pfx, CMS:stop2016, ATLAS:stop2016}, 
there are some unconstrained blind-spot regions for $m_\phi\gtrsim200$ GeV when $m_{\phi}\approx m_{\chi_2}$.

\subsection{The $S$}
\label{s.diphoton}
Another essential ingredient of the simplest hidden-sector implementation of WIMP-triggered baryogenesis is the neutral scalar mediator $S$. 
As we will discuss in this section, the detection of $S$ at the LHC will provide us with nontrivial pieces of information on the structure of the theory.
The connection of $S$ to the SM sector inevitably arises from the coupling $S\phi^\dag \phi$, 
which cannot be forbidden by any symmetry in the theory. 
At 1-loop level, this coupling generates couplings of $S$ to $gg$ and $\ga\ga$ through a loop of $\phi$.
It also necessarily generates couplings to $Z\ga$ and $ZZ$, but these are suppressed by the weak mixing angle. Compared to $S \to \ga\ga$, the rates for $S \to ZZ$ and $S \to Z\ga$ must be suppressed by $\tan^{4\!}\theta_w \simeq 9\%$ and $2\tan^{2\!}\theta_w \simeq 60\%$, respectively.
In this paper, we focus on the most dominant and cleanest channel, $gg \to S \to \ga\ga$, which could thus be the discovery channel of WIMP-triggered baryogenesis at the future LHC\@.

Before we proceed, let us comment on the variations of our model where $\phi$ has the gauge quantum numbers of $q$ or $d^\C$ instead of $u^\C$. 
If $\phi$ is $q$-like, $S$ would also acquire a coupling to $WW$.
However, such a case would correspond to a leptogenesis scenario and, as we noted in footnote~\ref{ftnote:leptogenesis}, that would require higher mass scales, pushing $S$ out of the LHC reach. 
If $\phi$ is $d^\C$-like, the loop-induced coupling of $S$ to $\ga\ga$, $\ga Z$, and $ZZ$ would be suppressed by a factor of $1/4$ because the hypercharge of $d^\C$ is half of that of $u^\C$. This would mean a suppression by a factor of $16$ in the $S$ production rate, again pushing $S$ out of the LHC reach.
Note that the vacuum stability constraint~\eq{kappa_bound} prohibits us from undoing this suppression by increasing $\ka$.
Therefore, 
since this section is about the LHC phenomenology of $S$, we do not consider the possibilities of $q$-like or $d^\C$-like $\phi$. 
Turning this around, if the $S$ is detected at the LHC, that will be a strong indication that $\phi$ is $u^\C$-like, not $q$- or $d^\C$-like, 
which is quite a nontrivial piece of information on the structure of the baryogenesis sector.

The detection of $S$ can also provide another interesting piece of information. Notice that the existence of 3 generations of $\phi$ to avoid excessive quark flavor violations leads to an enhancement by a factor of $3$ in the $gg \to S$ amplitude, and thus a factor of $9$ enhancement in the $S$ production rate, compared to the case with only one species of $\phi$, or the case with two generations of $\phi$ being much heavier as discussed in footnote~\ref{ftnote:decoupled_phi_12}.
Therefore, the detection of $S$ at the LHC will constitute a strong evidence that $\phi$ comes in a degenerate flavor multiplet and the flavor structure beyond the SM respects MFV\@.

\begin{figure}[t]
\includegraphics[width=\linewidth]{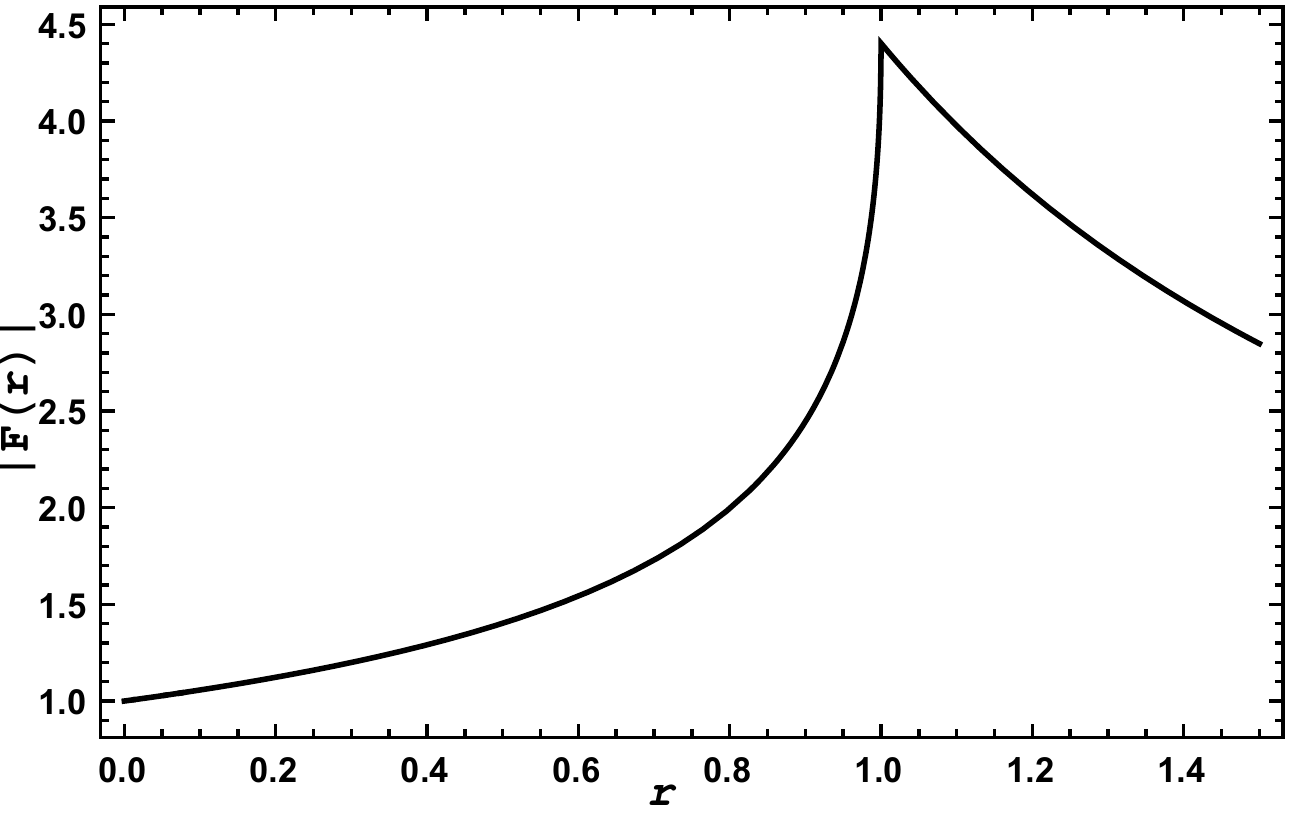} 
\caption{\label{f.F(r)}%
The function $F(r)$ defined in~\eq{F(r)}, which comes from the loop of $\phi$ in $S \to gg$ and $S \to \ga\ga$.}
\end{figure}

Now, let $\cM_{Sgg}$ be the amplitude for $gg \to S$. This $S$ may be off-shell if this process is part of a larger diagram.
At the 1-loop level, this amplitude is insensitive to the CP violations,
so the gauge, Lorentz, and CP invariances dictate that $\cM_{Sgg}$ have the following structure at 1-loop:
\beq
\cM_{Sgg} = 
\fr{\als}{4\pi \La_{g} (p_S^2)}
\lt[ (\ep_1 \dt q_2) (\ep_2 \dt q_1) -(\ep_1 \dt \ep_2) (q_1 \dt q_2) \rt] 
\,,\eql{MSgg} 
\eeq
where $q_i$ and $\ep_i$ ($i=1,2$) are the 4-momentum and polarization of the gluon $i$,
and it is understood that the colors of the two gluons are the same.
From an explicit calculation, the function $\La_g (p_S^2)$ is given by
\beq
\fr{1}{\La_g (p_S^2)}
= \fr{\kappa}{m_\phi^2} \cdot \fr12 \cdot 3 \cdot
\fr13 F\biggl( \fr{p_S^2}{4m_\phi^2} \biggr)
\,,\eql{1/La_g}
\eeq
where  
the factor of $1/2$ is from $\tr[T^a T^b] = \de^{ab}/2$, 
the factor of $3$ is due to the three generations of $\phi$,
and the factor of $1/3$ is introduced such that $F(0) = 1$,
where $F(r)$ is given by
\beq
F(r) 
&\equiv 
4! \int_0^1 \!\! \dd x \int_0^{1-x} \!\! \dd y \, 
\fr{xy}{1- 4rxy - \I 0^\text{\tiny +}}
\,.\eql{F(r)}
\eeq
In the range $0 < r < 1$, $F(r)$ is real and increases monotonically, starting from $F(0)=1$ and reaching $F(1) = 3(\pi^2-4) / 4 \simeq 4.4022$.
Above $r=1$, $F(r)$ acquires an imaginary part since the intermediate $\phi$ can be on-shell, with the magnitude $|F(r)|$ decreasing monotonically as $r$ increases.
(See \Fig{F(r)}.)
The analogous amplitude for $S \to \ga\ga$ is described by the amplitude 
of the same form as~\eq{MSgg} except that $\als$ and $\La_g$ are replaced 
by $\al$ and $\La_\ga$, respectively, where
\beq
\fr{1}{\La_\ga (p_S^2)}
= \fr{\kappa}{m_\phi^2} \cdot \fr49 \cdot
3 \cdot 3 \cdot \fr13
F\biggl( \fr{p_S^2}{4m_\phi^2} \biggr)
\,,\eql{1/La_ga}
\eeq
where the factor of $4/9$ is the square of the electric charge of $\phi$ 
and the two factors of $3$ are from the 3 colors and 3 generations of $\phi$.

From the amplitudes given above, we can immediately obtain the ratio of partial widths for $S \to \ga\ga$ and $S \to gg$:
\beq
\fr{\Ga_{S \to \ga\ga}}{\Ga_{S \to gg}}
=
\fr{\al^2}{8\als^2} \, \fr{\La_g^2 (m_S^2)}{\La_\ga^2 (m_S^2)} \ll 1
\,.\eql{gaga/gg}
\eeq
Therefore, if $m_S < 2m_\phi$, the width of $S$ is dominated by $S \to gg$ and given by
\beq
\Ga_S \simeq \fr{m_S}{8\pi} \lt( \fr{\als m_S}{4\pi \La_g} \rt)^{\!\! 2}
\,,
\eeq
which is very narrow.
If we instead have $m_S > 2m_\phi$, the $S$ can also decay to $\phi\phi$ with the partial width given by
\beq
\Ga_{S \to \phi\phi}
= \fr{3^2}{16\pi} \fr{\ka^2}{m_S} \sqrt{1 - \fr{4m_\phi^2}{m_S^2} }
\,.\eql{S-to-phiphi}
\eeq
Unless the phase space is nearly closed, this easily dominates over $S \to gg$ in the region of parameter space where $S \to \ga\ga$ may be observable in the first place, because $\ka$ must be large (i.e., $\sim \tev$ and hence $\sim m_S$) in such region.
Therefore, we expect that the chance of detecting $S$ in the $\ga\ga$ channel would significantly go down if $m_S > 2 m_\phi$.
In other words, the detection of $S$ in the $\ga\ga$ sample will strongly indicate the lower-bound on the $\phi$ mass, $m_\phi > m_S / 2$.

\begin{figure}[t]
\includegraphics[width=\linewidth]{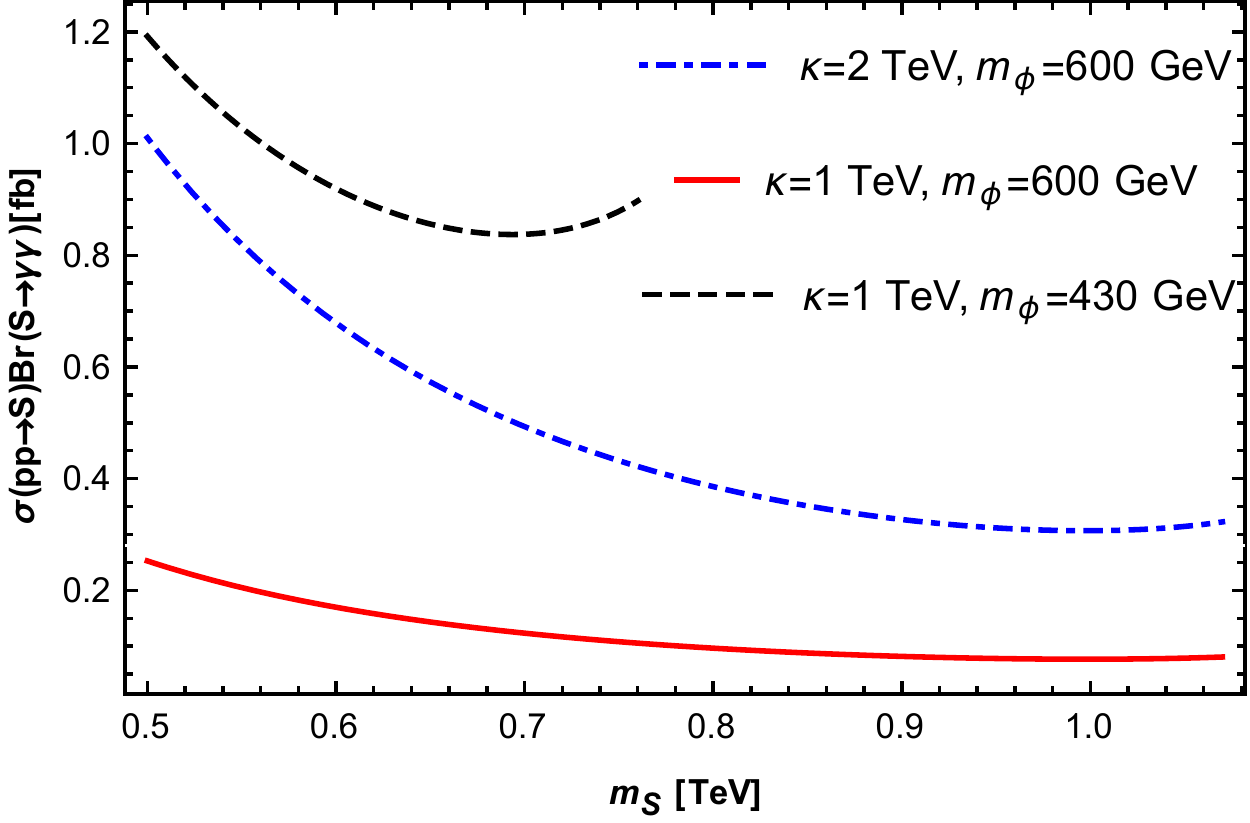} 
\caption{\label{f.diphoton_xsec}%
The diphoton production cross section via an $s$-channel $S$ as a function of $S$ at the LHC with $\sqrt{s} = 13\tev$.
See text for the reason why the black dashed line ends at $m_S = 760\gev$.
}
\end{figure}

\Fig{diphoton_xsec} shows the cross sections for diphoton production via $S$ at the 13-TeV LHC\@, using the MMHT2014 LO PDF \cite{Harland-Lang:2014zoa} evaluated at 
the factorization scale equal to $m_S$, for various benchmarks values of $\kappa$ and $m_\phi$. All benchmarks have $m_S < 2 m_\phi$ so that the diphoton branching fraction is not diluted any further by $S \to \phi\phi$ than it already is by $S \to gg$, since the purpose of this section is to study the prospect of the diphoton signal.
All values of $\ka$ and $m_S$ in the plot satisfy the vacuum stability condition~\eq{kappa_bound}, which in particular is the reason why the black dashed line in \Fig{diphoton_xsec} ends at about $m_S = 760\gev$.
The curves begin to go up toward larger values of $m_S$ because the values of $r$($=m_S^2 / 4m_\phi^2$) going into the function $F(r)$ are approaching 1 (see \Fig{F(r)}). 
The cusp of $F(r)$ at $r=1$ is an artifact of ignoring the width of $\phi$ in the calculation of $F(r)$. 
In other words, the expression~\eq{F(r)} becomes invalid if $r$ is too close to 1 to ignore the $\phi$ width.
However, the values of $r$ in \Fig{diphoton_xsec} never exceed $0.8$, which is safely far away from $1$ because the $\phi$ width is extremely small. 
(Recall that $\phi$ decays barely promptly and is generically long-lived.) 
The fact that $r < 0.8$ also justifies treating the intermediate $S$ on-shell and using the expression~\eq{gaga/gg} for the diphoton branching fraction, 
which in particular ignores the $S \to \phi_3 \bar{b} \bar{s}$ decay channel via an off-shell $\phi_3$.


Finally, although we have focused on $S \to \ga\ga$, there are other decay channels of $S$ induced by $\phi$ loop, i.e., $jj$ (from $gg$), $\gamma Z$, and $ZZ$.
The $WW$ mode is unlikely as we already discussed earlier in this subsection. 
Compared to $\ga\ga$, however, 
the $jj$ channel has much larger SM background and thus is much less clean,
while the $Z\ga$ and $ZZ$ have smaller production rates as we pointed out at the beginning of the subsection.
Therefore, 
we expect that the diphoton channel is the primary search channel.
All the curves in \Fig{diphoton_xsec} are below the 95\% CL upper bounds reported by the ATLAS~\cite{ATLAS:2016eeo} and CMS~\cite{Khachatryan:2016yec} collaborations 
but all within factors of a few at most. These are all benchmark curves, but it shows that the diphoton channel may well turn out to be the first collider signal of WIMP triggered baryogenesis.

\subsection{The WIMPs}
\label{s.WIMPs}

The WIMPs $\chi_{1,2}$ are, by definition, the central ingredients of the WIMP-triggered baryogenesis mechanism. The WIMP $\chi_1$ is meta-stable and decays to an SM quark and a $\phi$ with a large CP violation. The large CP violation is a consequence with the interference with a 1-loop diagram with virtual $\chi_2$ that has an $\cO(1)$ coupling to the same SM quark and $\phi$.
Therefore, at colliders, $\chi_1$ is either stable or long-lived, while $\chi_2$ decays promptly. Both decay to a jet and a $\phi$, so their decays are subsequently followed by the rich decay patterns of $\phi$ described in \Sec{phi_pheno}.
We thus see huge potential opportunities to probe the WIMP-triggered baryogenesis scenario at colliders through the productions of $\chi_1$ and $\chi_2$.

\subsubsection{Pair-production of $\chi_1$}
The meta-stable WIMP $\chi_1$ can be pair produced through $gg\rightarrow S\rightarrow \chi_1\chi_1$. 
If $m_{\chi_1} > m_S / 2$, the $s$-channel $S$ has to be off-shell 
and the $gg\rightarrow S\rightarrow \chi_1\chi_1$ cross section is proportional to $|y_1|^2\kappa^2$. 
This case is in principle very interesting because
the combination $|y_1|^2\kappa^2$ is fixed by the baryon abundance for any given $m_{\chi_1}$ as in~\eq{right-abundance}. 
Therefore, we \emph{predict} the $pp \rightarrow S^* \rightarrow \chi_1\chi_1$ production cross section as a function of $m_{\chi_1}$.
Unfortunately, the predicted cross sections falls far below an ab at the 13-TeV LHC, too small to be captured even by the high-luminosity LHC runs. 
It can be within the reach of the proposed next generation high luminosity 100 TeV $pp$ collider~\cite{Hinchliffe:2015qma}.
 
On the other hand, if $m_{\chi_1} < m_S / 2$, 
the $s$-channel $S$ in $gg\rightarrow S\rightarrow \chi_1\chi_1$ becomes on-shell.
Then, recalling the condition~\eq{mphi:upperbound}, 
the dominant $S$ decay channels should generically be $\chi_1 \chi_1$ and $\phi\phi$, 
since these processes occur at tree level while $S \to gg$ is 1-loop suppressed.
(We will discuss how the $\chi_1$ decays shortly.)
The $S$ production cross section is determined by $\kappa$ and $m_\phi$, 
while the branching fractions of $S$ into $\chi_1 \chi_1$ and $\phi\phi$ tell us about $y_1$ and $\ka$.
Then, we can test if the values of these parameters are consistent with baryogenesis
using the results of~\Sec{abundance}.
The partial width for $S \to \phi\phi$ is given already in~\eq{S-to-phiphi},
while that for $S \to \chi_1 \chi_1$ is given by
\beq
\Ga_{S \to \chi_1 \chi_1} 
= \frac{|y_1|^2 m_S}{16\pi} \,
v_{\chi_1} (\sin^{2\!} \de_1 + v_{\chi_1}^2 \!\cos^{2\!} \de_1 )
\,,\eql{S-to-chi1chi1}
\eeq
where $v_{\chi_1} \equiv \sqrt{1 - 4m_{\chi_1}^2 / m_S^2}$ is the speed of $\chi_1$ in the rest frame of the $S$.

\begin{figure}[]
\includegraphics[width=\linewidth]{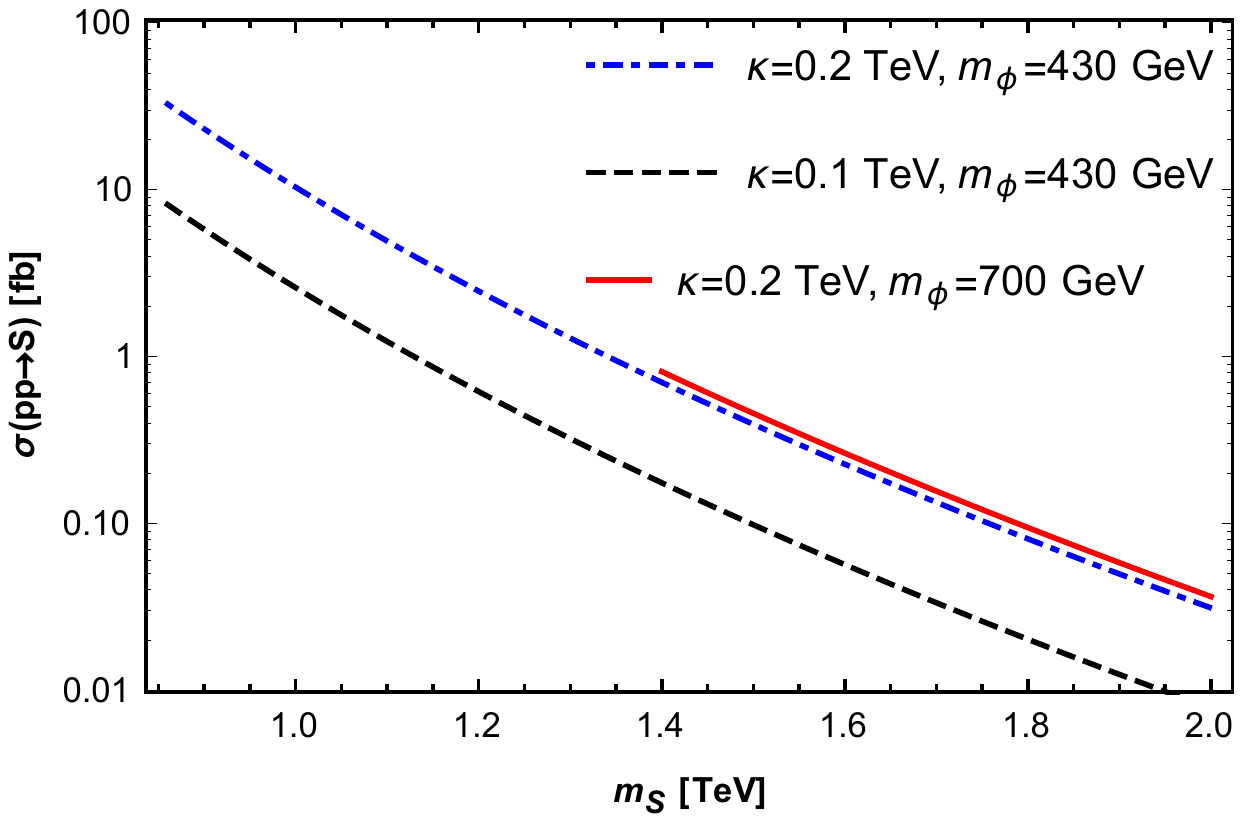} 
\caption{\label{f.S_onshell}%
The resonant $S$ production cross section as a function of $m_S$ at the LHC with $\sqrt{s} = 13\tev$.
The red solid line begins at $m_S = 1.4\tev$ because this plot is for the $m_S > 2m_{\chi_1}$ case and we also have the condition $m_{\chi_1} > m_\phi$ (Eq.~\eq{mphi:upperbound}).
Since $m_S > 2 m_{\chi_1, \phi}$, the leading $S$ decay channels are expected to be $\chi_1 \chi_1$ and $\phi\phi$, as discussed in text. See Eqs.~\eq{S-to-chi1chi1} and~\eq{S-to-phiphi} for the branching fractions of those modes.
}
\end{figure}

In \Fig{S_onshell}, the on-shell $S$ production cross sections are shown for various benchmark values of $\kappa$ and $m_\phi$. 
The very mild dependence on $m_\phi$ is due to an approximate accidental cancellation in Eq.~\eq{1/La_g} between $1/m_\phi^2$ and $F(m_S^2 / 4m_\phi^2)$ as we change $m_\phi$.
One sees in \Fig{F(r)} that increasing $m_\phi$ (thus decreasing $r$) increases $F(r)$ rather rapidly in the $r>1$ region.
\Fig{S_onshell} shows that the observation of $gg \to S \to \chi_1 \chi_1$ can be within the LHC reach if it has an $\cO(1)$ branching fraction. 
This will provide quite direct probes of our baryogenesis scenario, if $\chi_1$ can decay to a jet and a $\phi$ within the detectors. 

It is interesting to note that the diphoton signal of \Sec{diphoton} and the $\chi_1$ pair production are complementary to each other in many ways.
If the diphoton signal of \Sec{diphoton} is observed, \Fig{diphoton_xsec} will tell us that $\ka \sim \tev$. 
Combining this with the baryon abundance condition~\eq{right-abundance} will then imply that $|y_1|$ is small, making the branching fraction of $S$ to $\chi_1\chi_1$ is even smaller than that to $\ga\ga$.
On the other hand, if the $\chi_1$ production is observed, the condition~\eq{right-abundance} will imply a small $\ka$, making the observation of the diphoton signal unlikely.

Now, once $\chi_1$ is produced, the subsequent collider phenomenology crucially depends on how it decays because $\chi_1$ is meta-stable.
If $m_{\chi_1} < m_S / 2$ and $m_S \lsim 800\gev$, our analysis in \Sec{phi_pheno} points to Case 2-b, 
because we have $m_\phi < m_{\chi_1} \lsim 400\gev$ and the only case that allows such a light $\phi$ is Case 2-b.
Furthermore, the cosmologically determined range of $\chi_1^\PS$ lifetime~\eq{lambda1window} generically leads to a collider stable $\chi_1$ or a displaced $\chi_1^\PS$ decay.
The former would just appear as $\MET$.
The latter would appear as the production of two displaced vertices, each of which gives a $\phi$ plus a jet, or a $\phi$ plus a top.
The $\phi$ then decays to $\MET$ and practically unobservable soft jets, as discussed in Case 2-b of \Sec{phi_pheno}.

On the other hand, if $m_{\chi_1} > m_S / 2$ and/or $m_S \gsim 800\gev$, then
the variety of $\phi$ decay channels discussed in \Sec{phi_pheno} begins to open up.
After each of the pair produced $\chi_1$ decays with a (very) displaced vertex into an up-type quark (possibly a top) and a $\phi$, 
the $\phi$ can subsequently decay to two down-type quarks or an up-type quark plus a lighter $\chi_2$, promptly or displaced.

Although the existing displaced vertex searches at ATLAS and CMS can cover most of the event topologies from cascade decays of displaced $\chi_1$, we would like to point out in some cases 
a new dedicated trigger/analysis may be in demand. One specific example is where $\chi_1$ undergoes a displaced decay to a light jet (or a boosted top jet such that the muon trigger may not be efficient) and a $\phi$, then the $\phi$ subsequently decays to a invisible $\chi_2$ plus a soft jet (as in Case 2-b) at a secondary vertex that is further displaced relatively to the $\chi_1$ decay vertex.
Assisted by a possibly sizable Lorentz boost from $\chi_1$ decay, the jet associated with the $\chi_2$ vertex may not be as soft as in typical Case 2-b. 
The full event would thus consists of two sets of displaced and isolated ``emerging jets''~\cite{Schwaller:2015gea} macroscopically apart from each other (one from $\chi_1$ decay and the other from $\chi_2$ decay) yet connected by a track of $\phi$ (or rather, an R-hadron of $\phi$). Each $\chi_1$ decay also comes with $\MET$ from $\chi_2$ decay. The jet from $\chi_2$ decay may not be visible if $\chi_2$ does not have sufficient boost, in which case the (charged) R-hadron track would appear as a disappearing track.

Finally, we would like to make a brief remark that the recently proposed ``lifetime frontier'' detector MATHUSLA~\cite{Chou:2016lxi} (MAssive Timing Hodoscope for Ultra Stable neutraL pArticles) can greatly enhance sensitivity to long-lived particles such as $\chi_1$ at the high luminosity LHC\@. 
Ref.~\cite{Chou:2016lxi} also proposes a dedicated detector for a future 100 TeV collider, which can cover lifetimes as large as the limit allowed by BBN, $c\tau \sim 10^7$--$10^8$~m.

\subsubsection{Pair-production of $\chi_2$}
Although $\chi_2$ in our model does not directly trigger baryogenesis, it is in fact indispensable for generating a CP asymmetry necessary for baryogenesis. 
Recall that $\chi_2$ has an $\cO(1)$ coupling to an up-type quark and a $\phi$. 
This enables an appreciable tree-level pair production of $\chi_2$ from $u\bar{u}$ or $c\bar{c}$ through a $t$-channel $\phi$ exchange. 
($\chi_2$ can also be pair produced from $gg$ via an $s$-channel $S$, but the coupling of $gg$ to $S$ is loop suppressed, although the loop suppression may be partly countered by the large gluon PDF.\@) 
Numerical examples of $\chi_2$ pair-production rates are listed in TABLE \ref{table} .
\begin{table}[t]
\caption{\label{table} Pair production rates of $\chi_2$ from $t$-channel $\phi$ exchange with $\la_2 = 1$ for different values of $m_\phi$ and $m_{\chi_2}$ at the 13 TeV LHC\@ (simulated using the {\tt FeynRules 2.3}~\cite{Alloul:2013bka} and {\tt MadGraph 5}~\cite{Alwall:2014hca} packages).}
\centering
\begin{tabularx}{\columnwidth}{|X|X|X|}
\hline
$m_\phi ~(\gev)$ & $m_{\chi_2}~ (\gev)$ & $\sigma_{pp\to\chi_2\chi_2}~(\fb)$\\ \hline
600 & 400 & 32.7 \\ \hline
700 & 500 & 12.8 \\ \hline
700 & 900 & 1.2 \\ 
\hline
\end{tabularx}
\end{table}
If the diphoton signal of \Sec{diphoton} is observed with $m_S \lsim 800\gev$,
there are two possible scenarios corresponding to Case 1-a and Case 2-b of \Sec{phi_pheno}. In Case-1a, $\chi_2$ is heavier than $\phi$ so the pair-produced $\chi_2$'s promptly decay to a $\phi$ and an up-type quark,
where the up-type quark may be a top quark if the phase space is open. 
Note that the two $\phi$'s from the two $\chi_2$'s can be of different flavors,
and their charges do not have to be opposite. 
So, the $\chi_2$ pair production serves as a mechanism to produce two $\phi$'s with all possible combinations of charges and flavors together with additional $jj$, $tj$, $\bar{t}j$, $tt$, $t\bar{t}$, or $\bar{t}^{}\bar{t}$.
The $\phi$'s then subsequently decays to $bbjj$ promptly. 
Alternatively, in Case 2-b, $\phi$ is slightly heavier than $\chi_2$ and $\chi_2$ is collider stable. In this case, $\chi_2^\PS$ undergoes 3-body decay mediated by a barely off-shell $\phi$ as described in more detail in \Sec{phi_pheno}. 
On the other hand,
if the diphoton signal is not observed for $m_S \lsim 800\gev$,
the very rich $\phi$ decay modes discussed in \Sec{phi_pheno} can be realized, including possibly displaced multi-top/multi-bottom productions.

\begin{center}
*\hspace{3em}*\hspace{3em}*
\end{center}

To conclude the whole article, 
the WIMP-triggered baryogenesis mechanism not only provides a unified thermal WIMP origin of baryonic and dark components of matter 
but can also exhibit a rich collider phenomenology that allows us to probe the mechanism at the LHC,
even in the most pessimistic scenario that the WIMPs are completely neutral under the SM gauge group as it may be hinted by the null results of WIMP searches thus far. 
The LHC signals we have discussed include a clean diphoton resonance at the weak scale
and an array of other rich signatures that emerge from this WIMP-triggered baryogenesis mechanism, possibly displaced multi-bottom/multi-top productions, emerging jets, and (disappearing tracks of) R-hadrons.
We have also pointed out that di-nucleon decay provides us with a powerful probe of the mechanism.
Therefore, if this mechanism is indeed realized in nature, it is quite possible that we will be able to shed a bright, and first, light on one of the most fundamental questions in physics.\\

\begin{acknowledgments}
We thank Christian Reuschle, Brian Shuve and Daniel Stolarski for helpful discussions. AY is grateful to MadGraph and FeynRules teams for their technical support on MG5 and FeynRules 2.3. YC is supported by Perimeter Institute for Theoretical Physics, which is supported by the Government of Canada through
Industry Canada and by the Province of Ontario through the Ministry of
Research and Innovation. YC is also supported in part by the Maryland
Center for Fundamental Physics.
TO and AY are supported by the US Department of Energy under grant DE-SC0010102.
\end{acknowledgments}


\bibliography{refs.bib}{}

\bibliographystyle{apsrev4-1}

\end{document}